\documentclass{aip-cp}

\usepackage[numbers]{natbib}
\usepackage{rotating}
\usepackage{graphicx}
\usepackage{tikz}
\usepackage{pgfplots}
\usepgfplotslibrary{groupplots}
\pgfplotsset{compat=newest}

\begin{document}

% Title portion
\title{The Finite Difference Time Domain (FDTD) Method to Determine Energies and Wave Functions of Two-Electron Quantum Dot}

\author[aff1]{I Wayan Sudiarta\corref{cor1}}
\eaddress[url]{http://fisika.unram.ac.id/sudiarta}
\author[aff1]{Lily Maysari Angraini}

\affil[aff1]{Physics Study Program, Faculty of Matematics and Natural Sciences, University of Mataram, Mataram, NTB, Indonesia 83125}
\corresp[cor1]{wayan.sudiarta@unram.ac.id}

\maketitle

\begin{abstract}
The finite difference time domain (FDTD) method has been successfully applied to obtain energies and wave functions for two electrons in a quantum dot modeled by a three dimensional harmonic potential. The FDTD method uses the time-dependent Schrödinger equation (TDSE) in imaginary time. The TDSE is numerically solved with an initial random wave function and after enough simulation time, the wave function converges to the ground state wave function. The excited states are determined by using the same procedure for the ground state with additional constraints that the wave function must be orthogonal with all lower energy wave functions. The numerical results for energies and wave functions for different parameters of confinement potentials are given and compared with published results using other numerical methods. It is shown that the FDTD method gives accurate energies and wave functions.
\end{abstract}

% Head 1
\section{INTRODUCTION}

The finite difference time domain (FDTD) method is a general numerical method that can be applied to solve many physical problems such as electromagnetic wave simulations \cite{taflove}, acoustic wave simulations \cite{botteldooren1995} and quantum simulations \cite{sullivan2001, sullivan2002, sullivan2005, sudiarta2007}. The FDTD method has many advantages. Beside its good accuracy, it is also easy to code and very flexible for systems of any geometrical shapes. 

To solve the time-independence Schr\"odinger equation, the FDTD method can be devided into two methods: real-time FDTD (R-FDTD) method \cite{sullivan2001, sullivan2005} and imaginary-time FDTD (I-FDTD) method \cite{sudiarta2007, sudiarta2008, roy2014, sudiarta2016}. The R-FDTD method uses evolution of a wavefunction by the discretized time-dependence Schr\"odinger equation and Fourier transformation procedure to obtain eigen energies and wavefunctions. The I-FDTD method uses a diffusion equation to evolve an initial wavefunction and after a large interval of time the wavefunction approaches the ground state wavefunction. Advantanges of using I-FDTD method are (1) it only uses real variables compared to complex variables in the R-FDTD method for non magnetic case, (2) it does not require Fourier transformation and an absorbing boundary condition to truncate the computational domain.

In this paper, we apply the I-FDTD method to compute energies and wavefunctions of two electrons in a quantum dot. In the two-electrons system we need to descritize a six dimensional differential equation that requires very large number of grids. One approach to overcome this problem is to use an approximate model such as Hartree-Fock (HF) model \cite{sullivan2001}. The HF model does not contain correlation term which is important to simulate correlated system. Therefore solving full numerical Schr\"odinger equation is desired. In this paper, in order to get accurate results, an extrapolation procedure as given in \cite{crater1975} is done using FDTD results with various grid spacing parameters.

\section{THEORY}

In this paper, we briefly explain the FDTD method. A detail explanation of the FDTD method can be found in \cite{sudiarta2007}. To apply the FDTD method, we begin with the time-dependent Schr\"odinger equation (TDSE) for two electrons given by 
\begin{equation}
i\hbar \frac{\partial }{\partial t}\psi ({{\mathbf{r}}_{1}},{{\mathbf{r}}_{2}},t)= \hat{H}\psi ({{\mathbf{r}}_{1}},{{\mathbf{r}}_{2}},t) = \left[ -\sum\limits_{j=1}^{2}{\frac{{{\hbar }^{2}}}{2m}{{\nabla }_{j}}^{2}}+V({{\mathbf{r}}_{1}},{{\mathbf{r}}_{2}}) \right]\psi ({{\mathbf{r}}_{1}},{{\mathbf{r}}_{2}},t)
\label{eqn-tdse}
\end{equation}
\noindent where $m$ is the effective mass of electron and $V(\mathbf{r}_1,\mathbf{r}_2)$ is the potential given by

\begin{equation}
V(\mathbf{r}_1,\mathbf{r}_2) = V_{dot}(\mathbf{r}_1) + V_{dot}(\mathbf{r}_2) + \frac{e^2}{4\pi \epsilon_0 |\mathbf{r}_1 - \mathbf{r}_2|} 
\end{equation}
with an harmonic confining potential of a quantum dot $V_{dot}(\mathbf{r})= \frac{1}{2}m \left( \omega_x^2 x^2 + \omega_y^2 y^2 + \omega_z^2 z^2\right)$.

\noindent The solution of Eq.~(\ref{eqn-tdse}) can be expressed in the form of expansion of eigen functions  as     
\begin{equation}
\psi ({{\mathbf{r}}_{1}},{{\mathbf{r}}_{2}},t)=\sum\limits_{n=0}^{\infty }{{{c}_{n}}{{\phi }_{n}}({{\mathbf{r}}_{1}},{{\mathbf{r}}_{2}})\exp (-i{{E}_{n}}t/\hbar )}
\end{equation}

\noindent where $c_{n}$ is the expansion coefficient, the eigen functions $\phi_{n}(\mathbf{r}_{1}, \mathbf{r}_{2})$ and the energies $E_{n}$ are obtained by solving the time-independent Schr\"odinger equations (TISE), $\hat{H}\phi = E\phi$.

Beside the Schr\"odinger equation, the wavefunction $\phi_{n}(\mathbf{r}_{1},\mathbf{r}_{2})$ for electrons having the same spin must also satisfy the Pauli exclusion principle where the wave function must be anti-symmetric under exchange of two electrons \cite{bransden2000}, i.e.

\begin{equation}
{{\phi }_{n}}({{\mathbf{r}}_{1}},{{\mathbf{r}}_{2}})=-{{\phi }_{n}}({{\mathbf{r}}_{2}},{{\mathbf{r}}_{1}})			
\end{equation}

For simplicity in programming, in the remaining part of this paper we use atomic units with $\hbar = m =1$ and $e^2/4\pi\epsilon_0 = 1$. The first step in solving the TISE with the TDSE (Eq.~(\ref{eqn-tdse})) using the FDTD method is to transform Eq.~(\ref{eqn-tdse}) into a diffusion equation by substituting $\tau =it$ in Eq.~(\ref{eqn-tdse}). The resulting diffusion equation is 
\begin{equation}
\frac{\partial }{\partial \tau }\psi ({{\mathbf{r}}_{1}},{{\mathbf{r}}_{2}},\tau )=\frac{1}{2}\sum\limits_{j=1}^{2}{{{\nabla }_{j}}^{2}\psi ({{\mathbf{r}}_{1}},{{\mathbf{r}}_{2}},\tau )}-V({{\mathbf{r}}_{1}},{{\mathbf{r}}_{2}})\psi ({{\mathbf{r}}_{1}},{{\mathbf{r}}_{2}},\tau )		
\label{eqn-difusi}
\end{equation}
and the wavefunction is 
\begin{equation}
\psi ({{\mathbf{r}}_{1}},{{\mathbf{r}}_{2}},\tau )=\sum\limits_{n=0}^{\infty }{{{c}_{n}}{{\phi }_{n}}({{\mathbf{r}}_{1}},{{\mathbf{r}}_{2}},\tau )\exp (-{{E}_{n}}\tau )}
\label{eqn-psi}
\end{equation}.
It can be noted from Eq.~(\ref{eqn-psi}) that for a large imaginary time $\tau$, because of the factor $\exp(-E_n\tau)$, wave functions $\phi_n ({{\mathbf{r}}_{1}},{{\mathbf{r}}_{2}})$ with high energies diminish and the lowest energy state (or ground state) remains. Therefore after a large simulation time, the final wavefunction is 

\begin{equation}
 \lim_{\tau \to \infty} \psi ({{\mathbf{r}}_{1}},{{\mathbf{r}}_{2}},\tau )\approx [{{c}_{0}\exp (-{{E}_{0}}\tau )}]\ {{\phi }_{0}}({{\mathbf{r}}_{1}},{{\mathbf{r}}_{2}})
\end{equation}

\noindent where $[{{c}_{0}\exp (-{{E}_{0}}\tau )}]$ is just a constant factor which can be removed by normalizing the final wavefunction. Using an initial function $\psi ({{\mathbf{r}}_{1}},{{\mathbf{r}}_{2}},\tau =0)$, we can obtain the ground state wave function by a simulation using Eq.~(\ref{eqn-difusi}) for long enough interval of $\tau$ provided that the initial function contains the ground state or ${{c}_{0}}\ne 0$. Excited states can be obtained using same procedure provided that all lower energy wavefunctions have been removed from the initial function.   

After obtaining a wavefunction $\psi$, its energy can be computed by 
\begin{equation}
E = \frac{<\psi |\hat{H}|\psi >}{<\psi |\psi >}=\frac{\int{{{\psi }^{*}}\hat{H}\psi {{d}^{3}}{{r}_{1}}{{d}^{3}}{{r}_{2}}}}{\int{{{\left| \psi  \right|}^{2}}{{d}^{3}}{{r}_{1}}{{d}^{3}}{{r}_{2}}}}
\label{eqn-energy}
\end{equation}

\section{NUMERICAL METHOD}

For numerical computation, we use a notation ${{\psi }^{n}}({{i}_{1}},{{j}_{1}},{{k}_{1}},{{i}_{2}},{{j}_{2}},{{k}_{2}})=\psi ({{i}_{1}}\Delta x,{{j}_{1}}\Delta x,{{k}_{1}}\Delta x,{{i}_{2}}\Delta x,{{j}_{2}}\Delta x,{{k}_{2}}\Delta x,n\Delta \tau )$ (where $\Delta \tau $ and$\Delta {{x}_{{}}}$ are the temporal and spatial spacing. Following derivation in \cite{sudiarta2007}, it can be obtained that an explicit numerical iterative scheme for simulating evolution of a wavefunction is given by
\begin{equation}
\psi^{n+1}(i_1,j_1,k_1,i_2,j_2,k_2)=\alpha \psi^{n}(i_1,j_1,k_1,i_2,j_2,k_2)
+\beta \varphi^n(i_1,j_1,k_1,i_2,j_2,k_2)
\end{equation}
where $\varphi^n(i_1,j_1,k_1,i_2,j_2,k_2)$ is
\begin{equation}
\varphi^n(i_1,j_1,k_1,i_2,j_2,k_2) = \left[
\begin{array}{l}
\psi^{n}(i_1+1,j_1,k_1,i_2,j_2,k_2) + \psi^{n}(i_1-1,j_1,k_1,i_2,j_2,k_2)\\ 
+\psi^{n}(i_1,j_1+1,k_1,i_2,j_2,k_2) + \psi^{n}(i_1,j_1-1,k_1,i_2,j_2,k_2)\\
+\psi^{n}(i_1,j_1,k_1+1,i_2,j_2,k_2) + \psi^{n}(i_1,j_1,k_1-1,i_2,j_2,k_2)\\
+\psi^{n}(i_1,j_1,k_1,i_2+1,j_2,k_2) + \psi^{n}(i_1,j_1,k_1,i_2-1,j_2,k_2)\\
+\psi^{n}(i_1,j_1,k_1,i_2,j_2+1,k_2) + \psi^{n}(i_1,j_1,k_1,i_2,j_2-1,k_2)\\
+\psi^{n}(i_1,j_1,k_1,i_2,j_2,k_2+1) + \psi^{n}(i_1,j_1,k_1,i_2,j_2,k_2-1)\\
-12 \psi^{n}(i_1,j_1,k_1,i_2,j_2,k_2)
\end{array}
\right]
\end{equation}
and the coefficients $\alpha $ and $\beta $ are 

\begin{equation}
\alpha ({{i}_{1}},{{j}_{1}},{{k}_{1}},{{i}_{2}},{{j}_{2}},{{k}_{2}})=\frac{[1-\frac{\Delta \tau }{2}V({{i}_{1}},{{j}_{1}},{{k}_{1}},{{i}_{2}},{{j}_{2}},{{k}_{2}})]}{[1+\frac{\Delta t}{2}V({{i}_{1}},{{j}_{1}},{{k}_{1}},{{i}_{2}},{{j}_{2}},{{k}_{2}})]}
\end{equation}

\begin{equation}
\beta {{i}_{1}},{{j}_{1}},{{k}_{1}},{{i}_{2}},{{j}_{2}},{{k}_{2}}=\frac{\Delta \tau }{2(\Delta x)^{2}[1+\frac{\Delta t}{2}V({{i}_{1}},{{j}_{1}},{{k}_{1}},{{i}_{2}},{{j}_{2}},{{k}_{2}})]}
\end{equation}

In order to have a stable iteration, the time step $\Delta \tau $ must satisfy a stability condition given by
\begin{equation}
\Delta \tau \le (\Delta x)^{2}/6
\end{equation}.

In this paper to initiate the iteration, a random wavefunction is used as the initial wavefunction and the computational boundary is set to be zero or $\psi(\mathbf{r}_1,\mathbf{r}_2)_{boundary}=0$. The wavefunction for same spin electrons must also satisfy the Pauli exclusion principle which can be accomplished by updating the wavefunction at every step using  
\begin{equation}
\psi_{AS}({\mathbf{r}_{1}},{\mathbf{r}_{2}},\tau )=\frac{1}{2}[\psi ({\mathbf{r}_{1}},{\mathbf{r}_{2}},\tau )-\psi ({\mathbf{r}_{2}},{\mathbf{r}_{1}},\tau )]
\end{equation}

\noindent or in a discretized form given by

\begin{equation}
\psi^{n}_{AS}(i_1,j_1,k_1,i_2,j_2,k_2)=\frac{1}{2}[\psi^{n}(i_1,j_1,k_1,i_2,j_2,k_2)-\psi^{n}(i_2,j_2,k_2,i_1,j_1,k_1)]
\end{equation}

The energy is computed numerically by an approximation of Eq. ~(\ref{eqn-energy}) given by

\begin{equation}
E=\frac{1}{\sum \psi(i_1,\ldots,k_2)^{2}}\sum{\left\{V(i_1,\ldots, k_2)\psi {{(i_1,\ldots,k_2)}^{2}} -\psi (i_1,\ldots,k_2)\varphi(i_1,\ldots,k_2)/2\Delta {{x}^{2}} 
\right\}} 
\end{equation} 

\section{RESULTS AND DISCUSSIONS}

\subsection{Non Interacting Electrons In a Box}
To validate the FDTD method, we perform FDTD simulations for two non-interacting electrons in a cubic box with side length of one. The potential of the box is given by $V_{box}(\mathbf{r}) = 0$ inside the box ($|x|< 1/2$, $|y|< 1/2$, $|z|< 1/2$) and $V_{box}(\mathbf{r}_1, \mathbf{r}_2) = \infty$ outside the box. Numerical results for ground state and first excited state energies using various grid parameters are given in Table \ref{tbl-square}. It is noted that the numerical FDTD results approach exact energies as the grid spacing becomes smaller. 

The grid spacing is equal to the length of computation domain $L$ divided by number of grids $N$, that is $\Delta x = L/N$. The FDTD computation for two electrons requires $(N+1)^6$ grid points. This needs more than $40\times (N+1)^6$ bytes of computer memory when using double precision computation. Using a regular PC with four gigabites of memory can do computations with maximum number of grids per dimension $N = 17$. 

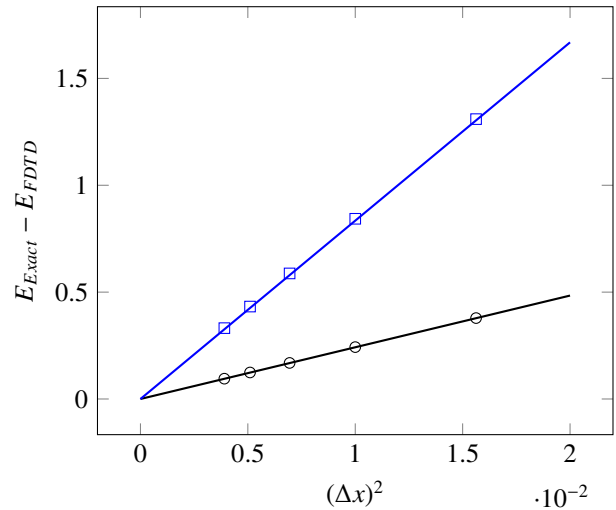
\begin{figure}[!h]
\begin{tikzpicture}
\begin{axis}[
  xlabel=$(\Delta x)^2$,
  ylabel=$E_{Exact} - E_{FDTD}$
  %,legend style={at={(0.5,-0.0)},anchor=west}
]
\addplot[color=black,only marks, mark=o]  table [x expr=\thisrowno{0}, y expr=\thisrowno{1}, col sep=space]{sqwell.dat}; 
%\addlegendentry{$f(x)$};
\addplot[solid, black,domain=0:0.02,samples=100,thick]{24.19457362*x};
%\addlegendentry{$\psi_1(x)$}
\addplot[color=blue,only marks, mark=square]  table [x expr=\thisrowno{0}, y expr=\thisrowno{2}, col sep=space]{sqwell.dat};
%\addlegendentry{$f(x)$};
\addplot[solid, blue,domain=0:0.02,samples=100,thick]{83.43347881*x};
%\addlegendentry{$\psi_2(x)$};
\end{axis}
\end{tikzpicture}
\caption{Numerical Errors in eigen energies as a function of $(\Delta x)^2$ for (1) ground state (circles) and (2) first excited state (squares). Lines are regression lines.}
\label{fig-box}
\end{figure}

To get accurate results we perform extrapolation using FDTD results for various spatial intervals $\Delta x$. Because the FDTD method uses central finite difference (CFD) scheme for the second derivative in the Schr\"odinger equation, the numerical errors of the CFD scheme are proportional to $(\Delta x)^2$. Numerical errors for ground state and excited state energies are plotted in Fig. \ref{fig-box}. The numerical errors are perfectly matched by a linear curve fitting. Therefore a fitting function used in this paper in the form of
\begin{equation}
E(\Delta x) = E - m (\Delta x)^2
\label{eqn-fitting}
\end{equation}    
is suitable for extrapolation. Other extrapolating functions such as Pad\'e approximating function and the Richardson method may also be used \cite{crater1975}. 

Ground state and excited state energies obtained by extrapolation are given in Table \ref{tbl-square}. It is shown clearly that a significant improvement is resulted when using this extrapolation method.     

\begin{figure}[!h]
\begin{tikzpicture}
\begin{axis}[
  xlabel=$x$,
  ylabel=$\psi(x)$
  %,legend style={at={(0.5,-0.0)},anchor=west}
]
\addplot[color=black,only marks, mark=*]  table [x expr=\thisrowno{0}, y expr=\thisrowno{1}, col sep=space]{psibox.dat}; 
%\addlegendentry{$f(x)$};
\addplot[solid, red,domain=0:1,samples=100,thick]{8*sin(deg(3.14159*x))};
\end{axis}
\end{tikzpicture}
\caption{Comparison of numerical ground state wavefunction (symbols: $\bullet$) with analytical wavefunction (line) for two non-interacting electrons in a box. $\psi(x) = \psi(x,0,0,0,0,0)$}
\label{fig-psibox}
\end{figure}
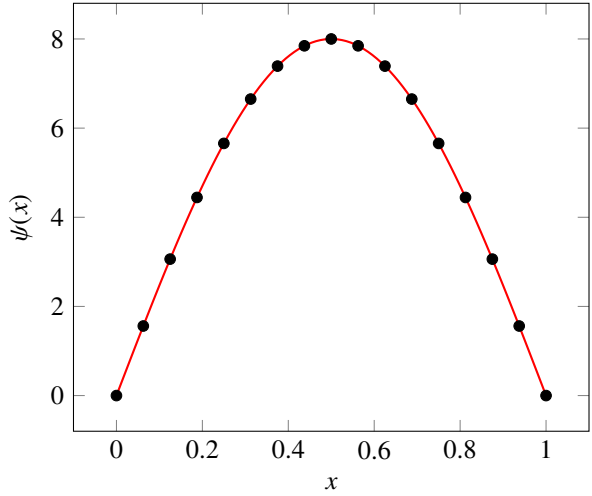

As an example of comparison between numerical and analytical wavefunctions are shown in Fig. \ref{fig-psibox} for values at the $x$ axis, that is the values of $\psi(x,0,0,0,0,0)$. The numerical wavefunction is in a good agreement with the analytical wavefunction.   
 
\begin{table}[!hbt]
\caption{\label{tbl-square} Numerical ground and first excited state energies with its errors for two electrons in a box (side length = 1) computed by the FDTD method using various number of grids $N$ or grid spacing $\Delta x$. The exact ground dan first excited state energies are $E_0 = 29.608813$ and $E_1 = 44.413220$. Numerical extrapolated energies are computed by extrapolating function Eq.~(\ref{eqn-fitting}.}
\tabcolsep7pt
\begin{tabular}{@{}rrrrrr}
\hline
N	&	$\Delta x$	&	$E_{FDTD,0}$	&	$E_{FDTD,1}$	&	$E_0 - E_{FDTD,0}$	&	$E_1 -E_{FDTD,1}$		\\
\hline
8	&	0.125000	&	29.230260	&	43.103716	&	0.378553	&	1.309503	\\
10	&	0.100000	&	29.366090	&	43.570042	&	0.242723	&	0.843177	\\
12	&	0.083333	&	29.440087	&	43.825748	&	0.168726	&	0.587471	\\
14	&	0.071429	&	29.484776	&	43.980749	&	0.124037	&	0.432470	\\
16	&	0.062500	&	29.513810	&	44.081681	&	0.095003	&	0.331538	\\
\multicolumn{2}{c}{Extrapolation (see text)}& 29.608196 & 44.406182& 0.000618 & 0.007038 \\
\hline
\end{tabular}
\end{table}
 
\subsection{Two Electrons in an Isotropic Quantum Dot}

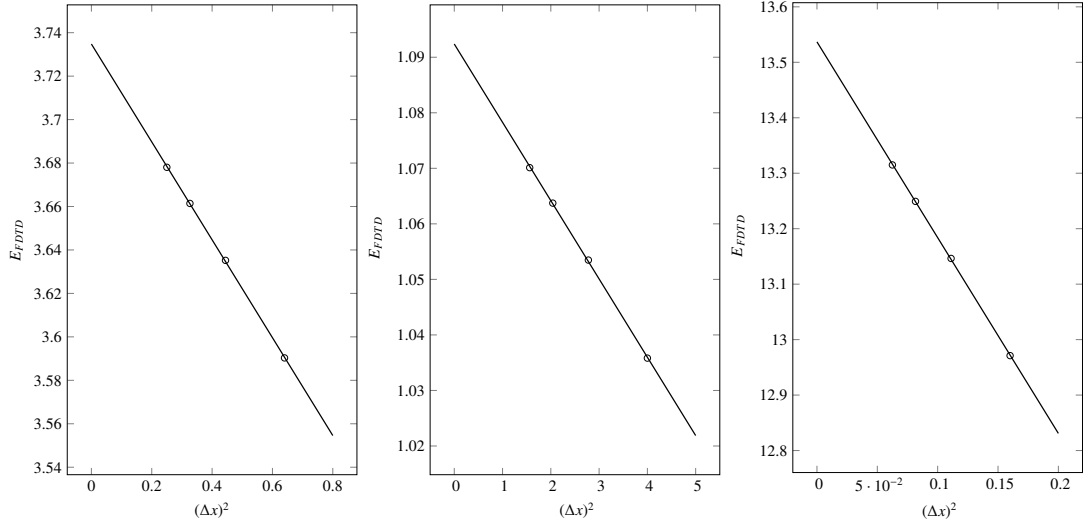
\begin{figure}[!h]
\newlength\figureheight 
\newlength\figurewidth 
\setlength\figureheight{12cm} 
\setlength\figurewidth{8cm}
\begin{tikzpicture}[scale = 0.60]
\begin{axis}[
width=\figurewidth,
height=\figureheight,
  xlabel=$(\Delta x)^2$,
  ylabel=$E_{FDTD}$
  %,legend style={at={(0.5,-0.0)},anchor=west}
]
\addplot[color=black,only marks, mark=o]  table [x expr=\thisrowno{0}, y expr=\thisrowno{1}, col sep=space]{qdot2.dat}; 
%\addlegendentry{$f(x)$};
\addplot[solid, black,domain=0:0.8,samples=100,thick]{-0.225257571*x + 3.734785302};
%\addlegendentry{$\psi_1(x)$}
\end{axis}
\end{tikzpicture}

\begin{tikzpicture}[scale = 0.60, font=\normalsize]
\begin{axis}[
width=\figurewidth,
height=\figureheight,
  xlabel=$(\Delta x)^2$,
  ylabel=$E_{FDTD}$
  %,legend style={at={(0.5,-0.0)},anchor=west}
]
\addplot[color=black,only marks, mark=o]  table [x expr=\thisrowno{2}, y expr=\thisrowno{3}, col sep=space]{qdot2.dat}; 
%\addlegendentry{$f(x)$};
\addplot[solid, black,domain=0:5,samples=100,thick]{-0.014100506*x + 1.092382419};
%\addlegendentry{$\psi_1(x)$}
\end{axis}
\end{tikzpicture}

\begin{tikzpicture}[scale = 0.60, font=\normalsize]
\begin{axis}[
width=\figurewidth,
height=\figureheight,
  xlabel=$(\Delta x)^2$,
  ylabel=$E_{FDTD}$
  %,legend style={at={(0.5,-0.0)},anchor=west}
]
\addplot[color=black,only marks, mark=o]  table [x expr=\thisrowno{4}, y expr=\thisrowno{5}, col sep=space]{qdot2.dat}; 
%\addlegendentry{$f(x)$};
\addplot[solid, black,domain=0:0.2,samples=100,thick]{-3.52984512*x + 13.53686
};
%\addlegendentry{$\psi_1(x)$}
\end{axis}
\end{tikzpicture}
\caption{Numerical eigen energies as a function of $(\Delta x)^2$ of two electrons in an isotropic quantum dot ($\omega_x = \omega_y = \omega_z = \omega$ for $\omega = 1$ (left figure), $\omega = 0.25$ (middle) and $\omega = 4$ (right). Lines are linear regression curves: (1) $E(\Delta x) =-0.225257571(\Delta x)^2 + 3.734785302$, (2) $E(\Delta x) = -0.014100506(\Delta x)^2 + 1.092382419$ and (3) $E(\Delta x) = -3.52984512(\Delta x)^2 + 13.53686$ .}
\label{fig-dot1}
\end{figure}

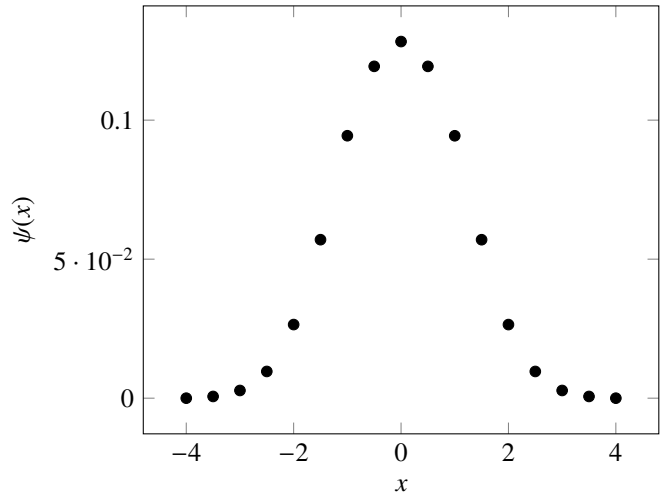
\begin{figure}[!h]
\begin{tikzpicture}
\begin{axis}[
  xlabel=$x$,
  ylabel=$\psi(x)$
  %,legend style={at={(0.5,-0.0)},anchor=west}
]
\addplot[color=black,only marks, mark=*]  table [x expr=\thisrowno{0}-4, y expr=\thisrowno{1}, col sep=space]{psidot.dat}; 
\end{axis}
\end{tikzpicture}
\caption{Numerical ground state wavefunction for two interacting electrons with different spin in an isotropic quantum dot ($\omega_x = \omega_y = \omega_z = 1$. $\psi(x) = \psi(x,0,0,0,0,0)$}
\label{fig-psidot}
\end{figure}

To demonstrate further application of the FDTD method, we consider two interacting electrons with different spins in an isotropic quantum dot ($\omega_x = \omega_y = \omega_z$ with three harmonic confining potentials ($\omega = 0.25, 1, 4$). Numerical results for ground state energies for four grid spacings are given in Fig. ~\ref{fig-dot1}. Similar to previous section, linear variation of energies as a function of $(\Delta x)^2$ are also shown for this case. After extrapolation procedure (see Fig. ~\ref{fig-dot1}), numerical ground state energies for the three confining potentials can be found in Table \ref{tbl-dot}. The FDTD results are in a good agreement with results of the discrete variable representation (DVR) method given in \cite{prudente2005}. A variation of numerical wavefunction of two electrons in quantum dot along x-axis is given in Fig.~\ref{fig-psidot}. It is expected due to harmonic confinement and singlet state, the maximum wavefunction is located in the middle of the quantum dot.   

\begin{table}[!hbt]
\caption{\label{tbl-dot} Numerical ground state energies computed using the FDTD method with extrapolation (see Fig.~\ref{fig-dot1}).}
\tabcolsep7pt
\begin{tabular}{@{}rrrr}
\hline
$\omega$	&	FDTD	&	ref \cite{prudente2005}\\
\hline
0.25	&	1.092382	&	1.089262\\
1	&	3.734785	&	3.730120\\
4	&	13.536860	&	13.523214\\
\hline
\end{tabular}
\end{table}

\section{CONCLUSION}
A numerical method known as the FDTD method to obtain eigen energies and wavefunctions for two electrons in a confining potential has been presented. The FDTD method for two electrons requires large amount of computer memory, therefore limited number of grids with large grid spacings are used. One method to overcome this limitation is by performing FDTD simulations with various grid spacings and then extrapolating the FDTD results. It has been shown that errors of the FDTD results are proportional to the square of the grid spacing. The FDTD method  with extrapolation gives accurate eigen energies and wavefunctions that have been validated by comparing with results from other analytical and numerical methods. 

\section{ACKNOWLEDGMENTS}
This work is partially supported by the University of Mataram under PNBP research scheme.  

%\nocite{*}
\bibliographystyle{aipnum-cp}
\bibliography{references}

\end{document}